\documentclass[a4paper,fleqn]{article}
\usepackage{graphics}
\newcommand{\be}{\begin{equation}}
\newcommand{\ee}{\end{equation}}
\newcommand{\ba}[1]{\begin{array}{#1}}
\newcommand{\ea}{\end{array}}
\newcommand{\n}[1]{\overline{#1}}
\newcommand{\bra}[1][\Psi_0]{\ensuremath{\langle #1 |}}
\newcommand{\ket}[1][\Psi_0]{\ensuremath{| #1 \rangle}}
\newcommand{\braket}[2]{\ensuremath{\langle #1 | #2 \rangle}}
\newcommand{\proj}[1]{\ensuremath{|#1\rangle\langle #1|}}
\newcommand{\neel}{N\'{e}el}
\newcommand{\trace}{{\rm tr~}}

\newcommand{\mas}{\renewcommand{\arraystretch}{1.3}}

\renewcommand{\phi}{\varphi}

\newcommand{\vertlaoii}{        %
\begin{minipage}{20mm}          %
\unitlength0.625mm              %
\begin{picture}(32,20)          %
\thinlines                      %
\put( 0,16){\line(1, 0){32}}    %
\put(16,16){\line(0,-1){16}}    %
\put(16,16){\circle*{4}}        %
\put( 5,16){\line(2, 1){6}}     %
\put( 5,16){\line(2,-1){6}}     %
\put(21,16){\line(2, 1){6}}     %
\put(21,16){\line(2,-1){6}}     %
\put(16,11){\line(-1,-2){3}}    %
\put(16,11){\line( 1,-2){3}}    %
\end{picture}                   %
\end{minipage}}                 %

\newcommand{\vertlaoio}{        %
\begin{minipage}{20mm}          %
\unitlength0.625mm              %
\begin{picture}(32,20)          %
\thinlines                      %
\put( 0,16){\line(1, 0){32}}    %
\put(16,16){\line(0,-1){16}}    %
\put(16,16){\circle*{4}}        %
\put( 5,16){\line(2, 1){6}}     %
\put( 5,16){\line(2,-1){6}}     %
\put(21,16){\line(2, 1){6}}     %
\put(21,16){\line(2,-1){6}}     %
\put(16, 5){\line(-1,2){3}}     %
\put(16, 5){\line( 1,2){3}}     %
\end{picture}                   %
\end{minipage}}                 %

\newcommand{\vertlaooi}{        %
\begin{minipage}{20mm}          %
\unitlength0.625mm              %
\begin{picture}(32,20)          %
\thinlines                      %
\put( 0,16){\line(1, 0){32}}    %
\put(16,16){\line(0,-1){16}}    %
\put(16,16){\circle*{4}}        %
\put( 5,16){\line(2, 1){6}}     %
\put( 5,16){\line(2,-1){6}}     %
\put(27,16){\line(-2, 1){6}}    %
\put(27,16){\line(-2,-1){6}}    %
\put(16,11){\line(-1,-2){3}}    %
\put(16,11){\line( 1,-2){3}}    %
\end{picture}                   %
\end{minipage}}                 %

\newcommand{\vertlaooo}{        %
\begin{minipage}{20mm}          %
\unitlength0.625mm              %
\begin{picture}(32,20)          %
\thinlines                      %
\put( 0,16){\line(1, 0){32}}    %
\put(16,16){\line(0,-1){16}}    %
\put(16,16){\circle*{4}}        %
\put( 5,16){\line(2, 1){6}}     %
\put( 5,16){\line(2,-1){6}}     %
\put(27,16){\line(-2, 1){6}}    %
\put(27,16){\line(-2,-1){6}}    %
\put(16, 5){\line(-1,2){3}}     %
\put(16, 5){\line( 1,2){3}}     %
\end{picture}                   %
\end{minipage}}                 %

\newcommand{\vertlaiii}{        %
\begin{minipage}{20mm}          %
\unitlength0.625mm              %
\begin{picture}(32,20)          %
\thinlines                      %
\put( 0,16){\line(1, 0){32}}    %
\put(16,16){\line(0,-1){16}}    %
\put(16,16){\circle*{4}}        %
\put(11,16){\line(-2, 1){6}}    %
\put(11,16){\line(-2,-1){6}}    %
\put(21,16){\line(2, 1){6}}     %
\put(21,16){\line(2,-1){6}}     %
\put(16,11){\line(-1,-2){3}}    %
\put(16,11){\line( 1,-2){3}}    %
\end{picture}                   %
\end{minipage}}                 %

\newcommand{\vertlaiio}{        %
\begin{minipage}{20mm}          %
\unitlength0.625mm              %
\begin{picture}(32,20)          %
\thinlines                      %
\put( 0,16){\line(1, 0){32}}    %
\put(16,16){\line(0,-1){16}}    %
\put(16,16){\circle*{4}}        %
\put(11,16){\line(-2, 1){6}}    %
\put(11,16){\line(-2,-1){6}}    %
\put(21,16){\line(2, 1){6}}     %
\put(21,16){\line(2,-1){6}}     %
\put(16, 5){\line(-1,2){3}}     %
\put(16, 5){\line( 1,2){3}}     %
\end{picture}                   %
\end{minipage}}                 %

\newcommand{\vertlaioi}{        %
\begin{minipage}{20mm}          %
\unitlength0.625mm              %
\begin{picture}(32,20)          %
\thinlines                      %
\put( 0,16){\line(1, 0){32}}    %
\put(16,16){\line(0,-1){16}}    %
\put(16,16){\circle*{4}}        %
\put(11,16){\line(-2, 1){6}}    %
\put(11,16){\line(-2,-1){6}}    %
\put(27,16){\line(-2, 1){6}}    %
\put(27,16){\line(-2,-1){6}}    %
\put(16,11){\line(-1,-2){3}}    %
\put(16,11){\line( 1,-2){3}}    %
\end{picture}                   %
\end{minipage}}                 %

\newcommand{\vertlaioo}{        %
\begin{minipage}{20mm}          %
\unitlength0.625mm              %
\begin{picture}(32,20)          %
\thinlines                      %
\put( 0,16){\line(1, 0){32}}    %
\put(16,16){\line(0,-1){16}}    %
\put(16,16){\circle*{4}}        %
\put(11,16){\line(-2, 1){6}}    %
\put(11,16){\line(-2,-1){6}}    %
\put(27,16){\line(-2, 1){6}}    %
\put(27,16){\line(-2,-1){6}}    %
\put(16, 5){\line(-1,2){3}}     %
\put(16, 5){\line( 1,2){3}}     %
\end{picture}                   %
\end{minipage}}                 %

\newcommand{\vertlbiii}{\raisebox{0.5\depth}{\rotatebox{180}{\vertlaiii}}}
\newcommand{\vertlbiio}{\raisebox{0.5\depth}{\rotatebox{180}{\vertlaiio}}}
\newcommand{\vertlbioi}{\raisebox{0.5\depth}{\rotatebox{180}{\vertlaoii}}}
\newcommand{\vertlbioo}{\raisebox{0.5\depth}{\rotatebox{180}{\vertlaoio}}}
\newcommand{\vertlboii}{\raisebox{0.5\depth}{\rotatebox{180}{\vertlaioi}}}
\newcommand{\vertlboio}{\raisebox{0.5\depth}{\rotatebox{180}{\vertlaioo}}}
\newcommand{\vertlbooi}{\raisebox{0.5\depth}{\rotatebox{180}{\vertlaooi}}}
\newcommand{\vertlbooo}{\raisebox{0.5\depth}{\rotatebox{180}{\vertlaooo}}}

\newcommand{\vvertlaonoio}{     %
\begin{minipage}{30mm}          %
\unitlength0.625mm              %
\begin{picture}(48,20)          %
\thinlines                      %
\put( 0,16){\line(1, 0){48}}    %
\put(16,16){\line(0,-1){16}}    %
\put(32,16){\line(0,-1){16}}    %
\put(16,16){\circle*{4}}        %
\put(32,16){\circle*{4}}        %
\put( 5,16){\line(2, 1){6}}     %
\put( 5,16){\line(2,-1){6}}     %
\put(43,16){\line(-2, 1){6}}    %
\put(43,16){\line(-2,-1){6}}    %
\put(16,11){\line(-1,-2){3}}    %
\put(16,11){\line( 1,-2){3}}    %
\put(32, 5){\line(-1,2){3}}     %
\put(32, 5){\line( 1,2){3}}     %
\end{picture}                   %
\end{minipage}}                 %

\newcommand{\vvertlbmmmmn}{     %
\begin{minipage}{20mm}          %
\unitlength0.625mm              %
\begin{picture}(32,26)          %
\thinlines                      %
\put( 0,21){\line(1, 0){32}}    %
\put( 0, 5){\line(1, 0){32}}    %
\put(16,21){\line(0,-1){16}}    %
\put(16,21){\circle*{4}}        %
\put(16, 5){\circle*{4}}        %
\put( 3,23){$\mu_1$}            %
\put( 3, 7){$\mu_2$}            %
\put(24,23){$\mu_3$}            %
\put(24, 7){$\mu_4$}            %
\end{picture}                   %
\end{minipage}}                 %

\newcommand{\vvertlbMMMMn}{     %
\begin{minipage}{20mm}          %
\unitlength0.625mm              %
\begin{picture}(32,26)          %
\thicklines
\put( 0,21){\line(1, 0){32}}    %
\put( 0, 5){\line(1, 0){32}}    %
\put(16,21){\line(0,-1){16}}    %
\put(16,21){\circle*{4}}        %
\put(16, 5){\circle*{4}}        %
\put( 3,23){$\nu_1$}            %
\put( 3, 7){$\nu_2$}            %
\put(24,23){$\nu_3$}            %
\put(24, 7){$\nu_4$}            %
\end{picture}                   %
\end{minipage}}                 %

\newcommand{\vertlt}{           %
\begin{minipage}{20mm}          %
\unitlength0.625mm              %
\begin{picture}(32,26)          %
\thicklines                     %
\put( 0,20){\line(1, 0){32}}    %
\put( 0, 4){\line(1, 0){32}}    %
\put(17,22){\line(0,-1){18}}    %
\thinlines                      %
\put( 0,22){\line(1, 0){32}}    %
\put( 0, 6){\line(1, 0){32}}    %
\put(15,22){\line(0,-1){18}}    %
\put(16,21){\circle{4}}         %
\put(16, 5){\circle{4}}         %
\put( 3,24){$\mu_1$}            %
\put( 3, 8){$\mu_2$}            %
\put( 3,16){$\nu_1$}            %
\put( 3, 0){$\nu_2$}            %
\put(24,24){$\mu_3$}            %
\put(24, 8){$\mu_4$}            %
\put(24,16){$\nu_3$}            %
\put(24, 0){$\nu_4$}            %
\end{picture}                   %
\end{minipage}}                 %

\title{Optimum ground states for spin-$\frac{3}{2}$ ladders \\ with two legs}
\author{H.~Niggemann \and J.~Zittartz}
\date{\small Institut f\"ur Theoretische Physik, Z\"ulpicher Str. 77,
             D-50937 K\"oln}
\begin{document}
\maketitle
\begin{abstract}
We construct the exact ground state for an antiferromagnetic
spin-$\frac{3}{2}$ model on the two-leg ladder as an
{\em optimum ground state}. The ground state contains a discrete
parameter $\sigma=\pm 1$ and a continuous parameter $a$ which controls
$z$-axis anisotropy. For most values of $a$ the global ground state is
unique. It has vanishing sublattice magnetization and exponentially
decaying correlation functions. By using the transfer matrix
technique, we calculate exactly the fluctuations of the magnetization,
the nearest-neighbour correlation, and the longitudinal correlation
length as functions of the parameters. \\
\\
{\em Dedicated to Prof.~H.~Horner on the occasion of his 60th birthday.}
\end{abstract}

\renewcommand{\thefootnote}{}
\footnotetext{Work performed within the research program of the
Sonderforschungsbereich 341, K\"{o}ln-Aachen-J\"{u}lich}
\renewcommand{\thefootnote}{\arabic{footnote}}

\section{Introduction}
The investigation of quantum spin ladders is a very active field of
condensed matter physics. Reports concerning experimental realizations
of such systems are contributed frequently, see for instance
\cite{hnk}--\cite{fy}. A review can be found in \cite{dr}. 
From a theoretical point of view, spin ladders have been investigated
by using exact diagonalization \cite{f}, density matrix
renormalization group (DMRG) methods \cite{ls,kt}, bosonization
\cite{scp}, and various other techniques,
see e.g.\ \cite{ds}--\cite{bkmn} and references therein.

Recently, {\sc Kolezhuk} and {\sc Mikeska} \cite{km} presented a set
of {\em matrix product ground states} (MPG) for special isotropic
spin-$\frac{1}{2}$ ladder models. To each rung of the ladder, a matrix
is assigned, which has local spin states for the corresponding rung as
its elements. The global state is given by the product of these
matrices, in which the matrix elements are multiplied via the
tensorial product in spin space. If the model parameters obey the
conditions given in \cite{km}, the resulting global state is a so
called {\em optimum ground state}, i.e.\ it is not only a ground state
of the global Hamiltonian, but also of every local interaction operator.

In the present work we construct a one-parametric set of optimum
ground states for antiferromagnetic spin-$\frac{3}{2}$ ladders with
two legs. The local interaction is identical to the one on the
hexagonal lattice in our previous paper \cite{hexag32}. In contrast to
\cite{km}, the global ground state is given in terms of a
{\em vertex state model}, which is a generalization of the MPG
approach. This allows to specify the local spin states for every
lattice site instead for complete rungs as required by the MPG
approach. Properties of the ground state are calculated by using the
{\em transfer matrix technique}, which is explained in
Appendix~A.

\section{The model}
Consider a ladder of length $L$ with two legs and periodic boundary
conditions. Each lattice site is occupied by a spin-$\frac{3}{2}$. The
global Hamiltonian
\be
\label{h}
H=\sum_{i=1}^{L} \left[\, h_{i,i+1} + h_{i',i'+1} + h_{i,i'} \,\right]
\ee
contains only nearest neighbour interactions. If the ladder is
visualized horizontally, the index $i$ ($i'$) denotes the upper
(lower) spin on rung number~$i$. All local interactions are equal,
they only act on different spin pairs, i.e.\ the system is completely
homogeneous.

The local interaction operator is the same as in the
spin-$\frac{3}{2}$ model on the hexagonal lattice presented in our
previous work \cite{hexag32}. It is given in terms of projectors onto
its eigenstates:
\be
\label{hij}
\mas
\ba{lcl}
h_{ij}& = &
  \lambda_3 \,\left(\, \proj{v_3} + \proj{v_{-3}} \,\right)\, + \\
&&\lambda_2^{-\sigma} \,\left(\, \proj{v_2^{-\sigma}} +
                                 \proj{v_{-2}^{-\sigma}} \,\right)\, + \\
&&\lambda_{12}^+ \,\left(\, \proj{v_{12}^+} + \proj{v_{-12}^+} \,\right)\, + \\
&&\lambda_{02}^{-\sigma} \, \proj{v_{02}^{-\sigma}} \, ,
\ea\ee
where
\be
\label{lexcited}
\mas
\ba{lcl}
\ket[v_3] &=& \ket[33] \\
\ket[v_{-3}] &=& \ket[\n{33}] \\
\ket[v_2^{-\sigma}] &=& \ket[31]-\sigma\ket[13] \\
\ket[v_{-2}^{-\sigma}] &=& \ket[\n{31}]-\sigma\ket[\n{13}] \\
\ket[v_{12}^+] &=& a \ket[11]
    - \,\left(\, \ket[3\n{1}]+\ket[\n{1}3] \,\right)\, \\
\ket[v_{-12}^+] &=& a \ket[\n{11}]
    - \,\left(\, \ket[\n{3}1]+\ket[1\n{3}] \,\right)\, \\
\ket[v_{02}^{-\sigma}] &=& \sigma a^2 
    \,\left(\, \ket[1\n{1}]-\sigma\ket[\n{1}1] \,\right)\, -
    \,\left(\, \ket[3\n{3}]-\sigma\ket[\n{3}3] \,\right)\, .
\ea
\ee
$\lambda_3,\lambda_2^{-\sigma},\lambda_{12}^+,\lambda_{02}^{-\sigma}$
are positive real numbers, $a$ is real, and $\sigma=\pm 1$. The
canonical spin-$\frac{3}{2}$ basis states are denoted as
\be
\mas
\ba{lcl@{\hspace{2cm}}lcl}
S^z \ket[3]     &=&  \frac{3}{2} \ket[3] &
S^z \ket[\n{3}] &=& -\frac{3}{2} \ket[\n{3}] \\
S^z \ket[1]     &=&  \frac{1}{2} \ket[1] &
S^z \ket[\n{1}] &=& -\frac{1}{2} \ket[\n{1}] \, .
\ea
\ee
Note that (\ref{hij}) has the following properties:
\begin{enumerate}
\item It has rotational symmetry in the $xy$-plane of spin space,
i.e.\ it commutes with the local magnetization operator $S^z_i+S^z_j$.
\item It is parity invariant, i.e.\ it commutes with the operator
$P_{ij}$, which interchanges the spins at lattice sites $i$ and $j$.
\item It has spin-flip symmetry, i.e.\ it is invariant under the
transformation $S^z\to -S^z$.
\item Its lowest eigenvalue is zero, i.e.\ $h_{ij}$ is positive
semi-definite.
\end{enumerate}
The Hamiltonian contains 5 continuous parameters, namely 4
$\lambda$-parameters plus $a$. This includes a trivial scale.
The two-spin states (\ref{lexcited}) are the excited local eigenstates
of $h_{ij}$, the remaining 9 eigenstates are local ground states,
i.e.\ the corresponding eigenvalue is zero.

Since the global Hamiltonian (\ref{h}) is a sum of positive
semi-definite operators, zero is also a lower bound of the global
ground state energy $E_0$. In the next section a global eigenstate
corresponding to eigenvalue zero is constructed, which must therefore
be the global ground state.

For $a=-\sqrt{3}$ and $\sigma=-1$, the $\lambda$-parameters can be
chosen so that (\ref{lexcited}) are eigenstates of
$({\bf S}_i+{\bf S}_j)^2$. In this {\em isotropic case}, the local
interaction (\ref{hij}) has complete $SO(3)$ symmetry in spin space
and can be written as
\be
h_{ij} =             {\bf S}_i \cdot {\bf S}_j 
  + \frac{116}{243}( {\bf S}_i \cdot {\bf S}_j )^2
  +  \frac{16}{243}( {\bf S}_i \cdot {\bf S}_j )^3
  +  \frac{55}{108} \, ,
\ee
which simply projects onto all states with
$({\bf S}_i+{\bf S}_j)^2=3(3+1)$.

\section{The global ground state}
To each lattice site we assign a set of {\em vertices} with binary
bond variables which are denoted as arrows. The values of these
vertices are single-spin states at the corresponding lattice site. On
the upper row
\be
\label{vertupper}
\renewcommand{\arraystretch}{4}
\ba{rcr@{\qquad}rcr}
\vertlaooo&:&     a \ket[3]     &
\vertlaiii&:&     a \ket[\n{3}] \\
\vertlaooi&:&       \ket[1]     &
\vertlaiio&:&       \ket[\n{1}] \\
\vertlaioo&:&\sigma \ket[1]     &
\vertlaoii&:&\sigma \ket[\n{1}] \\
\vertlaoio&:&       \ket[1]     &
\vertlaioi&:&       \ket[\n{1}] \makebox[0em][l]{\, .}
\ea\ee
On the lower row
\be
\label{vertlower}
\renewcommand{\arraystretch}{4}
\ba{rcr@{\qquad}rcr}
\vertlbooo&:&       a \ket[3]     &
\vertlbiii&:&\sigma a \ket[\n{3}] \\
\vertlbooi&:&\sigma   \ket[1]     &
\vertlbiio&:&         \ket[\n{1}] \\
\vertlbioo&:&\sigma   \ket[1]     &
\vertlboii&:&         \ket[\n{1}] \\
\vertlboio&:&         \ket[1]     &
\vertlbioi&:&\sigma   \ket[\n{1}] \makebox[0em][l]{\, .}
\ea\ee
The parameters $a$ and $\sigma$ are the same as in (\ref{lexcited}).

In order to construct the global ground state from these local spin
states, it is necessary to define the concatenation of the vertices
(\ref{vertupper}) and (\ref{vertlower}), respectively. This is similar
to classical vertex models of statistical physics, but the generic
product of numbers is replaced by the tensorial product in spin space:
\be
\renewcommand{\arraystretch}{4}
\ba{rcl}
\vvertlaonoio &=& \vertlaoii \otimes \vertlaooo \\
              &+& \vertlaooi \otimes \vertlaioo \, ,
\ea
\ee
i.e.\ the interior bond between the two vertices is summed out.
Concatenations between adjacent vertices on the lower row and between
upper and lower vertex on the same rung are completely analogous. The
resulting cluster represents a state in the Hilbert space of two
neighbouring spins.

A global state $\ket$ can be constructed by successively attaching
vertices to the cluster until the ladder has the desired length. This
process is associative, i.e.\ the order in which the vertices are
concatenated does not matter. As periodic boundary conditions are
imposed, the `free' bonds at the first and last rung of the ladder
have to be summed out. Because of the similarity to classical vertex
models, such a global state is called a {\em vertex state model}
\cite{hexag32,nz}. These are direct generalizations of the well-known
{\em matrix product ground states} for spin chains \cite{ksz,nz}. If
$\ket[\phi_i^{\mu_1 \mu_2 \mu_3}]$ denotes the vertex at site $i$ with
arrow variables $\mu_1$, $\mu_2$, and $\mu_3$ (cf.\ definitions
(\ref{vertupper}) and (\ref{vertlower})), a formal expression for the
ground state is given by
\be
\ket = \sum_{\lbrace \mu \rbrace} \; \prod^{\otimes}_i \;
\ket[\phi_i^{\mu_1 \mu_2 \mu_3}] \, .
\ee
The sum is over all arrow configurations on the bonds.

In order to show that the resulting global state $\ket$ is the ground
state of the Hamiltonian (\ref{h}), we collect all two-spin state
which are generated by all possible concatenations of two
vertices\footnote{Common prefactors have been omitted.}:
\be
\label{lground}
\mas
\ba{l@{\hspace{2cm}}l}
\ket[31] + \sigma \ket[13] &
\ket[\n{31}] + \sigma \ket[\n{13}] \\
\ket[11] + a \ket[3\n{1}] &
\ket[\n{11}] + a \ket[\n{3}1] \\
\ket[11] + a \ket[\n{1}3] &
\ket[\n{11}] + a \ket[1\n{3}] \\
\ket[1\n{1}] + \sigma a^2 \ket[3\n{3}] &
\ket[\n{1}1] + \sigma a^2 \ket[\n{3}3] \\
\ket[1\n{1}] + \sigma \ket[\n{1}1] \, . & 
\ea
\ee
It is easy to check that each of these 9 two-spin states is
perpendicular to all local excited states (\ref{lexcited}), i.e.\ they
are annihilated by the local interaction operator (\ref{hij}). In
other words, (\ref{lground}) are the local ground states of $h_{ij}$.

From the construction of the vertex state model it is clear that any
projection of $\ket$ onto the Hilbert space of two neighbouring spins
is a linear combination of the two-spin states
(\ref{lground}). Therefore $\ket$ is annihilated by all local
interaction operators,
\be
h_{ij} \ket = 0
\ee
for all nearest neighbours $i$ and $j$. Hence
\be
H \ket = 0 \, .
\ee
As explained above, zero is a lower bound for the ground state energy
$E_0$, therefore $\ket$ must be the ground state of $H$. This type of
global ground state, which simultaneously minimizes all local
interaction operators, is called an {\em optimum ground state}
\cite{hexag32}--\cite{nz}, since $E_0$ takes the lowest possible value.

\section{Properties of the ground state}
The first interesting expectation value is the magnetization of a
single spin. As can be seen from the vertices (\ref{vertupper}) and
(\ref{vertlower}), a spin flip $S^z\to -S^z$ is equivalent to a flip
of all arrows on the bonds, which leaves the global state invariant.
Therefore the single-spin magnetization $\langle S_i^z\rangle_{\Psi_0}$
must vanish. In this sense the global ground state is antiferromagnetic.

For $a\to\infty$, the structure of the global ground state becomes
very simple. In this limit, the ground state is completely dominated
by the vertices
\be
\renewcommand{\arraystretch}{4}
\ba{rcr@{\qquad}rcr}
\vertlaooo&:&       a \ket[3]     &
\vertlaiii&:&       a \ket[\n{3}] \\
\vertlbooo&:&       a \ket[3]     &
\vertlbiii&:&\sigma a \ket[\n{3}] \makebox[0cm]{\hspace{1em} .}
\ea\ee
Clearly, there are only two different possibilities to concatenate
these four vertices to form a global state. These two possibilities
correspond to the two different \neel\ states on the ladder. Thus in
the limit $a\to\infty$, the global ground state is simply given by%
\footnote{The global prefactor $a^{2L}$ has been dropped.}
\be
\ket = \ket[\mbox{\neel}_1] + \sigma^L \, \ket[\mbox{\neel}_2] \; .
\ee

The {\em transfer matrix technique} can be used to calculate
expectation values for arbitrary values of the parameters $a$ and
$\sigma$. Appendix~A explains how to apply this method to the
two-leg ladder. For the present model, the eigensystem of the transfer
matrix can be obtained exactly, but has a very complicated form. For
this reason, explicit formulae for the expectation values have been
omitted. Instead, the results for the thermodynamic limit are plotted
as a function of the parameter $a$. The dependence on $\sigma$ drops
out in all expectation values, which are calculated in this work.

\begin{figure}[t]
\renewcommand{\tabcolsep}{1mm}
\begin{tabular}{rc}
\raisebox{2.5cm}{$\langle (\widetilde{S}^z_i)^2\rangle_{\infty}$} &
\resizebox{7cm}{!}{\includegraphics{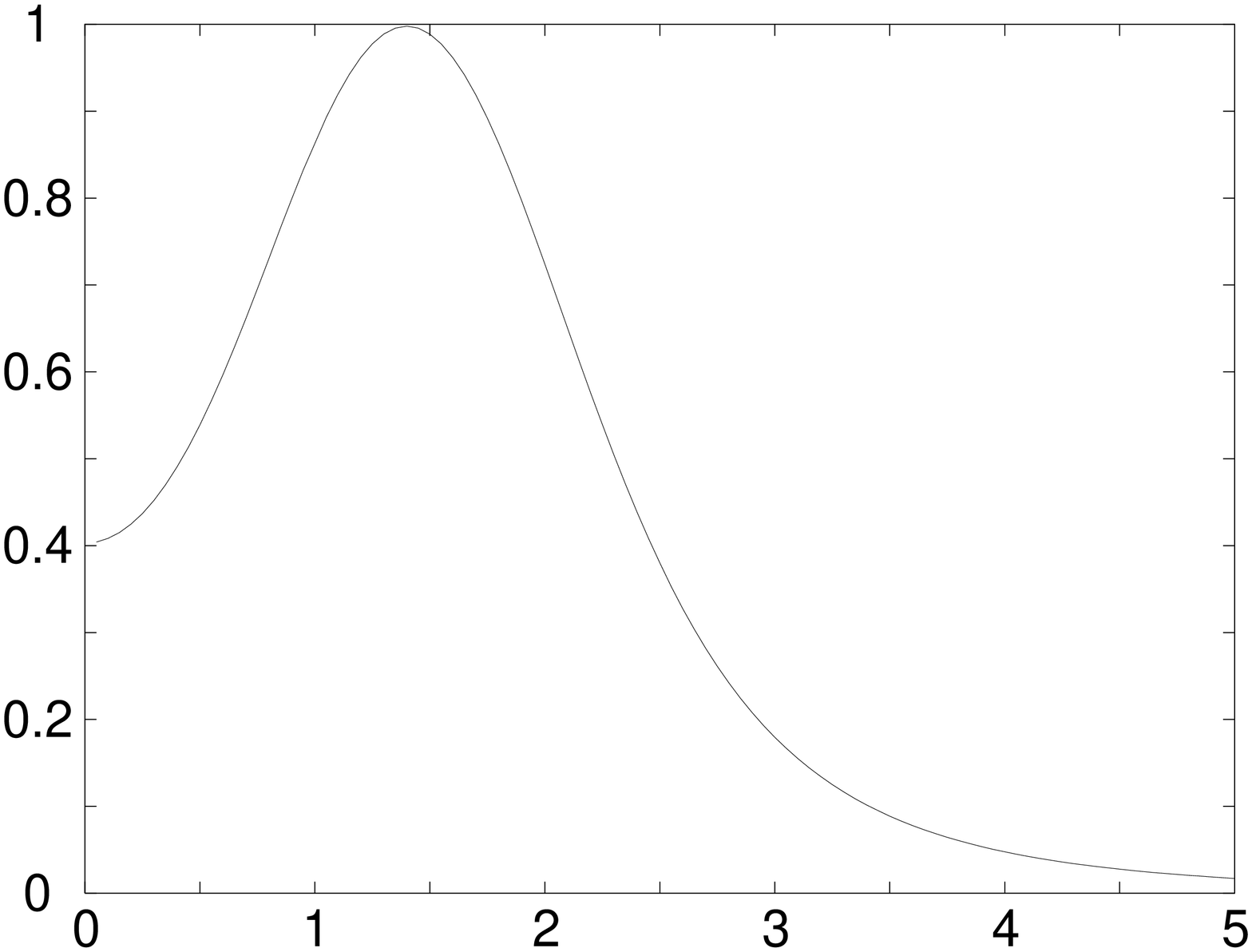}} \\
& $a$
\end{tabular}
\caption{Fluctuations of the pair magnetization as a function of the
parameter $a$}
\label{figuresz2}
\end{figure}

Consider the combined magnetization operator of an elementary cell at
rung number~$i$,
\be
\widetilde{S}^z_i = S^z_{i} + S^z_{i'} \; ,
\ee
where the index $i$/$i'$ again denotes the upper/lower spin on the rung,
respectively. Of course, $\langle\widetilde{S}^z_i\rangle_{\infty}$
vanishes. Its fluctuations can be calculated by using
equation~(\ref{singleexpectinf}). The result is shown in
figure~\ref{figuresz2}.
The maximum at $a\approx 1.4$ can be understood as follows: It is
clear from the vertices (\ref{vertupper}) and (\ref{vertlower}), that
for $a=0$ the system contains only the single-spin states $\ket[1]$
and $\ket[\n{1}]$. With increasing values of $a$, $\ket[3]$ and
$\ket[\n{3}]$ states are mixed in, so the fluctuations become
larger. Finally, for $a\to\infty$, the global ground state is simply a
superposition of both possible \neel\ states, so the system is
frozen. Therefore, the fluctuations must vanish in this limit.

\begin{figure}[t]
\renewcommand{\tabcolsep}{1mm}
\begin{tabular}{rc}
\raisebox{2.5cm}{$\langle \widetilde{S}^z_i \widetilde{S}^z_{i+1}
                  \rangle_{\infty}$} &
\resizebox{7cm}{!}{\includegraphics{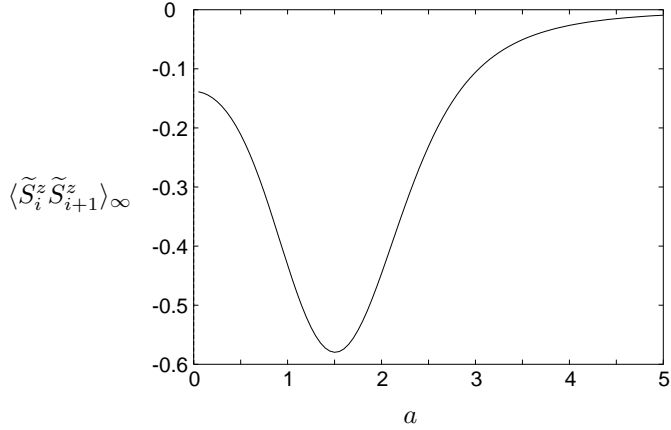}} \\
& $a$
\end{tabular}
\caption{Correlation between adjacent rungs as a function of $a$}
\label{figureszsz}
\end{figure}

Correlations between two different rungs can be obtained from
equation~(\ref{doubleexpectinf}). The nearest neighbour correlation,
$\langle \widetilde{S}^z_i \widetilde{S}^z_{i+1} \rangle_{\infty}$, is
plotted as a function of $a$ in figure~\ref{figureszsz}.
The shape is very similar to the one in figure~\ref{figuresz2}. Note
that the correlation is antiferromagnetic for all values of $a$. The
vanishing in the limit $a\to\infty$ becomes clear from
\be
\mas
\ba{rcl}
\langle \widetilde{S}^z_i \widetilde{S}^z_{i+1} \rangle_{\infty} &=&
   \langle S^z_{i}  S^z_{i+1}  \rangle_{\infty}
 + \langle S^z_{i'} S^z_{i+1'} \rangle_{\infty} \\
&+&\langle S^z_{i}  S^z_{i+1'} \rangle_{\infty}
 + \langle S^z_{i'} S^z_{i+1}  \rangle_{\infty} \; .
\ea\ee
The first two terms approach $-\frac{9}{4}$ in the \neel\ limit
$a\to\infty$, but the last two terms converge to $+\frac{9}{4}$, since
the spin operators act on the same sublattice. The minimum at
$a\approx 1.5$ is very close to the maximum of the fluctuations
(cf.\ figure~\ref{figuresz2}).

\begin{figure}[t]
\renewcommand{\tabcolsep}{1mm}
\begin{tabular}{rc}
\raisebox{2.5cm}{$\xi_l^{-1}$} &
\resizebox{7cm}{!}{\includegraphics{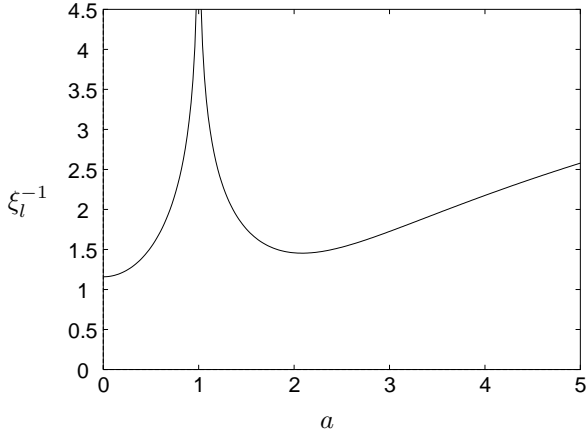}} \\
& $a$
\end{tabular}
\caption{Inverse longitudinal correlation length as a function of $a$}
\label{figurexil}
\end{figure}

As a function of the distance $r$, the correlation between
$\widetilde{S}^z_1$ and $\widetilde{S}^z_{r}$ decays exponentially,
i.e.\ $\langle \widetilde{S}^z_1 \widetilde{S}^z_{r} \rangle_{\infty}
\propto \exp(-r/\xi_l)$. Figure~\ref{figurexil} shows the
$a$-dependence of the corresponding inverse correlation length
$\xi_l^{-1}$.
Note the divergence at $a=1$. At this point, the weight of {\em all}
non-vanishing vertices of the corresponding classical vertex model
(cf.\ Appendix~A) is 1. This corresponds to an infinite
temperature, if the vertex weight is interpreted as a `Boltzmann
weight'. The correlation length remains finite for all values of $a$,
so the system is never critical.

For $a\neq 0$, the constructed optimum ground state $\ket$ on the
ladder is the {\em only} ground state of the global Hamiltonian
(\ref{h}). The proof can be carried out by induction according to
system size and is omitted here. In the special case $a=0$, the global
ground state degeneracy grows exponentially with system size.

\section{Summary and outlook}
We have constructed an {\em optimum ground state} for a 5-parametric
spin-$\frac{3}{2}$ model on the two-leg ladder. The Hamiltonian is
completely homogeneous and contains only nearest neighbour couplings.
The local interaction operator exhibits $S^z$-conservation and parity
invariance. No external magnetic field is applied. For special values
of the parameters, the system has complete $SO(3)$ symmetry.

Optimum ground states are not only ground states of the global
Hamiltonian, but also of every local interaction operator. For the
present model, the ground state is given in terms of a
{\em vertex state model}. A set of vertices, which have single-spin
states as their values, is assigned to each lattice site.
Concatenating these vertices on the two-leg ladder yields the global
ground state. Vertex state models are straightforward generalizations
of the well-known {\em matrix product ground states} for spin-chains.
The ground state contains a continuous parameter $a$ and a discrete
parameter $\sigma=\pm 1$.

Properties of the ground state have been calculated by using the
transfer matrix technique. The sublattice magnetization and the total
magnetization vanish, so the ground state is antiferromagnetic.
Two-point correlations along the ladder decay exponentially. The
fluctuations of the sublattice magnetization, the nearest neighbour
correlation, and the longitudinal correlation length have been
determined as a function of the parameter $a$. As the correlation
length is always finite, the system is never critical. Except for
$a=0$, the global ground state is unique.

For other spin-$\frac{3}{2}$ models on the two-leg ladder, e.g.\ with
generic Heisenberg interaction, the constructed ground state could
serve as a variational ground state. In this scenario, the anisotropy
parameter $a$ plays the role of a variational parameter. Optimum
ground states can also be constructed for more sophisticated ladder
models, e.g.\ with next-to-nearest neighbour interaction and
inhomogeneous couplings.

\appendix
\renewcommand{\theequation}{\Alph{section}.\arabic{equation}}
\renewcommand{\thesection}{Appendix \Alph{section}:}

\section{The transfer matrix technique}
\setcounter{equation}{0}
Consider a local observable $A_i$, which acts on the spins
on rung number~$i$. Its expectation value in the ground state is
defined as
\be
\label{singleexpect}
\langle A_i\rangle_{\Psi_0} =
\frac{\bra A_i \ket}{\braket{\Psi_0}{\Psi_0}} \, .
\ee
The denominator can be interpreted as two identical vertex state
models on top of each other. The upper (lower) one represents the
bra- (ket-) vector. Since the vertices at a given lattice site generate
only local spin states, the inner product can be taken separately at
each lattice site. The result is a {\em classical} vertex model,
i.e.\ the `weights' of the vertices are numbers, but each vertex
emanates {\em two} bonds in each direction. One set of bonds carries
the arrow variables of the bra-vector, the other one carries those of
the ket-vector. Therefore the inner product $\braket{\Psi_0}{\Psi_0}$
can be constructed from the $L$-fold product of the {\em transfer matrix}
\be
\label{transfer}
T_{(\mu_1,\mu_2,\nu_1,\nu_2)}^{(\mu_3,\mu_4,\nu_3,\nu_4)}=
\vertlt = \left\langle \; \vvertlbmmmmn \; \right| \left. \,
                          \vvertlbMMMMn \; \right\rangle \, .
\ee
It has 4 incoming ($\mu_1,\mu_2,\nu_1,\nu_2$) and 4 outgoing indices
($\mu_3,\mu_4,\nu_3,\nu_4$). Each index represents a binary bond
variable, so $T$ is a $16\times 16$-matrix. As we are dealing with
periodic boundary conditions, the trace must be used to obtain the
inner product
\be
\label{denominator}
\braket{\Psi_0}{\Psi_0} = \trace T^L \, .
\ee

The numerator of (\ref{singleexpect}) can also be interpreted as a
classical vertex model. It differs from the one for the denominator
only at rung number~$i$, where the observable $A_i$ acts
non-trivially. Therefore the numerator also has a matrix product
representation, but the $i$-th factor is the
{\em associated transfer matrix} $T(A)$ instead of $T$:
\be
\label{numerator}
\bra A_i\ket = \trace \left[\, T^{i-1} \, T(A) \, T^{L-i} \,\right]
             = \trace \left[\, T(A) \, T^{L-1} \,\right] \, .
\ee
$T(A)$ is obtained from definition (\ref{transfer}) by inserting $A_i$
between the bra- and the ket-vector on the r.h.s. of the equation.
Note that (\ref{numerator}) is independent of $i$, as $\ket$ has
perfect translational invariance. Since $T$ is symmetric, its
eigenvalues $\chi_k$ and its normalized eigenvectors $\ket[u_k]$ can
be used to calculate the $L$-fold matrix products. This yields
\be
\label{singleexpectsum}
\langle A_i \rangle_{\Psi_0} =
\frac{\sum_k \bra[u_k]T(A)\ket[u_k]\chi_k^{L-1}}{\sum_k \chi_k^L}
\ee
for the expectation value of $A_i$. In the thermodynamic limit
$L\to\infty$, which is the most interesting case, only the largest
eigenvalue $\chi_{\rm max}$ survives. This simplifies
(\ref{singleexpectsum}) to
\be
\label{singleexpectinf}
\langle A_i \rangle_{\infty} =
\frac{\bra[u_{\rm max}]T(A)\ket[u_{\rm max}]}{\chi_{\rm max}} \, .
\ee

The above consideration can be easily extended to two-point correlation
functions. In the thermodynamic limit the corresponding formula is
\be
\label{doubleexpectinf}
\langle A_1 B_r \rangle_{\infty} = \frac{1}{\chi_{\rm max}^2}
\sum_k \bra[u_{\rm max}]T(A)\ket[u_k]
       \bra[u_k]T(B)\ket[u_{\rm max}]
\left(\frac{\chi_k}{\chi_{\rm max}}\right)^{r-2} .
\ee

If the following mapping of matrix indices to arrow configurations is
used, the transfer matrix (\ref{transfer}) takes a block-diagonal form:
\be
\ba{r|cccc}
\mbox{index} &
\ba{c}\mbox{upper bra}\\ \mu_1/\mu_3 \ea &
\ba{c}\mbox{upper ket}\\ \nu_1/\nu_3 \ea &
\ba{c}\mbox{lower bra}\\ \mu_2/\mu_4 \ea &
\ba{c}\mbox{lower ket}\\ \nu_2/\nu_4 \ea \\
\hline
 1 & \leftarrow  & \leftarrow  & \leftarrow  & \leftarrow  \\
 2 & \rightarrow & \rightarrow & \rightarrow & \rightarrow \\
 3 & \leftarrow  & \leftarrow  & \rightarrow & \rightarrow \\
 4 & \rightarrow & \rightarrow & \leftarrow  & \leftarrow  \\
 5 & \leftarrow  & \rightarrow & \rightarrow & \leftarrow  \\
 6 & \rightarrow & \leftarrow  & \leftarrow  & \rightarrow \\
\hline
 7 & \leftarrow  & \leftarrow  & \leftarrow  & \rightarrow \\
 8 & \rightarrow & \rightarrow & \leftarrow  & \rightarrow \\
 9 & \leftarrow  & \rightarrow & \leftarrow  & \leftarrow  \\
10 & \leftarrow  & \rightarrow & \rightarrow & \rightarrow \\
\hline
11 & \leftarrow  & \leftarrow  & \rightarrow & \leftarrow  \\
12 & \rightarrow & \rightarrow & \rightarrow & \leftarrow  \\
13 & \rightarrow & \leftarrow  & \leftarrow  & \leftarrow  \\
14 & \rightarrow & \leftarrow  & \rightarrow & \rightarrow \\
\hline
15 & \leftarrow  & \rightarrow & \leftarrow  & \rightarrow \\
\hline
16 & \rightarrow & \leftarrow  & \rightarrow & \leftarrow 
\ea\ee
The first block is the $6\times 6$-matrix
\be
\label{tmsub1}
T_{1\to 6}=\left(\ba{cccccc}
2      & 2a^2   & 1+a^2  & 1+a^2  & \sigma & \sigma \\
2a^2   & 2      & 1+a^2  & 1+a^2  & \sigma & \sigma \\
1+a^2  & 1+a^2  & 2      & 1+a^4  & \sigma & \sigma \\
1+a^2  & 1+a^2  & 1+a^4  & 2      & \sigma & \sigma \\
\sigma & \sigma & \sigma & \sigma & 2      & 0      \\
\sigma & \sigma & \sigma & \sigma & 0      & 2
\ea\right) .
\ee
This submatrix contains the leading eigenvalue of the whole transfer
matrix for all values of the parameters.

The $4\times 4$-blocks for indices $7\to 10$ and $11\to 14$ are
identical. Both are given by
\be
\label{tmsub2}
\ba{l}
T_{7\to 10}=T_{11\to 14}= \\
\\
\left(\ba{cccc}
2\sigma           & \sigma(1\!+\!a^2) & 1 & 1 \\
\sigma(1\!+\!a^2) & 2\sigma           & 1 & 1 \\
1 & 1 & 2\sigma           & \sigma(1\!+\!a^2) \\
1 & 1 & \sigma(1\!+\!a^2) & 2\sigma
\ea\right) .
\ea\ee
The remaining 2-dimensional subspace of indices $15$ and $16$ is
already diagonal:
\be
\chi_{15}=\chi_{16}=2
\ee
Since both submatrices (\ref{tmsub1}) and (\ref{tmsub2}) are
symmetric, an orthogonal eigenbasis of the complete transfer matrix
exists. However, exact expressions for the eigensystem are very
complicated, so they have been omitted.

\end{document}